\documentstyle[preprint,aps,eqsecnum]{revtex}
\tightenlines
\begin{document}

\title{THE INTRINSIC DENSITY MATRICES OF NUCLEAR SHELL MODEL }
\author{A.Deveikis, G.Kamuntavi{\v c}ius}
\address{Vytautas Magnus University, Kaunas 3000, Lithuania}
\date{\today}
\maketitle

\begin{abstract}
     A new method for the calculation of shell model intrinsic density
     matrices, defined as two--particle density matrices integrated
     over centrum of mass position vector of two last particles and
     accompanied with isospin variables, has been developed.
     Produced intrinsic density matrices are completely antisymmetric,
     translational invariant and do not employ a group theoretical
     classification of antisymmetric states.
     They are devoted for exact realistic density matrix expansion within
     the framework of Reduced Hamiltonian method.
     The procedures on a base of precise arithmetic for intrinsic
     density matrices calculation that involve any numerical diagonalization
     and orthogonalization were developed and implemented in computer code.
\end{abstract}

\section{INTRODUCTION}

Last time discovering of a exotic nuclei emphasize the deficiency of
usual nuclei description methods based on central field approximation.
The reason relay upon week exotic nuclei binding energy due to oversaturation
of neutrons or protons.
To describe such the systems large--space shell model expansions have to be
considered, since the ordinary shell model is unable to manage deep
nucleon--nucleon correlations.
However shell model functions are implicitly redundant with many times
encountering a numerous of the same intrinsic states.
The reason is that excited state shell model wave function is representable
as linear combination of the products of intrinsic wave function and centrum
of mass (c.m.) wave function and only the first term in the series
corresponding to the c.m. in ground state is needful.
Every remaining term contains the one of excited intrinsic wave functions
that was encountered in the first term of expansion of less excited state
shell model wave function.
The first attempt to solve this problem was the translationally
invariant shell model \cite{1,2,3}, preliminary invoked to antisymmetryze
the wave-function depending on intrinsic coordinates, leads only to
sophisticated exercising in group theory which do not end in available
results.

We propose a simple and effective procedures comprising simplicity of
ordinary shell model and requirement of translational invariancy for
wave-function calculation. The method relies on the number of statements
simplifying the problem under consideration.

At first, despite the harmonic oscillator potential do not depend on
spin and isospin coordinates, we refuse of orbital, spin and isospin spaces
separation in model wave-functions. Hence we loose an advantage of using
the group theoretical methods for many-particle antisymmetrical states
classification. Nevertheless, on the other hand, rejection of precise
labelling of model wave-functions by quantum numbers of higher order groups,
such as seniority, leads to simplification of the very
antisymmetrization procedure. In this approach many-particle antisymmetrical
states are characterized only by well defined set of quantum numbers:
number of oscillator quanta $E$, angular momentum $J$, parity $\mit\Pi$,
isospin $T$ and only one additional integer quantum number
${\mit\Gamma} = 1,2, \ldots r$ necessary for unambiguous enumeration of the
states. Here $r$ is rank of the corresponding antisymmetrization operator
matrix. Developed on this basis computational procedures
allow unrestricted configuration space to be taken into account \cite{4,5}.

Second simplification refer to the 'spurious' states. Description of atomic
nuclei can be carried out within the framework of ordinary shell model
using intrinsic wave-function expansion in terms of shell model functions.
The coefficients of this expansion can be obtained diagonalizing c.m.
Hamiltonian in the shell model basis.
Intrinsic motion of nucleons would be represented by subspace of c.m.
Hamiltonian eigenvectors  corresponding to minimal eigenvalue equal $3\over 2$.

At last, simplification of the A-particle system description
is using two-particle density matrices instead of wave-functions.
Since all operators of observables including
intrinsic Hamiltonian and root-mean-square (r.m.s.) radius operators are
symmetric ones, their expectation values calculations
do not require the complete wave-function to be involved.
Moreover two-particle density matrix is still redundant in its variables.
We can use more simple quantity as density matrix integrated over c.m.
position vector of two last particles, as so-called intracule \cite{6}.
In case of light nuclei intracule have to describe A-particle system with
not excited c.m. and must be accompanied with isospin variables \cite{4}.
Such intracule we call intrinsic density matrix.
Goal is that using intrinsic density matrices we have to deal only with
$7+7$ orbital, spin and isospin variables, instead of wave-functions
containing $5A-3$ variables.

\section{DEVELOPMENT OF THE METHOD}

In the shell model it is assumed that free nucleons are moving in
self--consistent central field. It can be approximated by an isotropic
harmonic oscillator potential. The many--particle Hamiltonian used in
shell model is

\begin{eqnarray}
  \label{H_r}
  H = \sum_{i=1}^A
    \left\{
     - \frac{\hbar^2}{2m}\Delta_i +
     {1\over 2}m\omega^2 r_i^2
    \right\}
    = \sum_{i=1}^A
    h(\vec r_i)
\end{eqnarray}

Here $m$ denotes the nucleon mass, $\vec r_i$ the i-th nucleon radius vector,
$\omega$ the angular frequency and $h(\vec r_i)$ the single-particle hamiltonian.
Usually in the shell model only one eigenfunction of many--particle
Hamiltonian~(\ref{H_r}) is taken into account. Hence the consideration
is restricted by only one configuration (so--called ground configuration)
characterized by minimal total oscillator quantum number.
The consistent consideration could be achieved only on complete basis of
eigenfunctions of Hamiltonian~(\ref{H_r}).
To construct any antisymmetrical eigenfunction of many--particle
$H$~(\ref{H_r}) characterized with a good quantum numbers it is appropriate
to define eigenfunctions of single-particle $h(\vec r_i)$ in a j-j coupled
representation as

 \begin{eqnarray}
  \label{singl_f_r}
   \begin{array}{r}
     \psi_{eljtm_jm_t}(x)
     =
     R_{el}(r)
     \{
       Y_l(\widehat{\vec r})
       \otimes
       \alpha_{1/2}(\vec\sigma)
     \}_{jm_j}\;
     \alpha_{1/2m_t}(\vec\tau)
   \end{array}
 \end{eqnarray}

Here $R_{el}(r)$ is a radial function,
$Y_{lm}(\widehat{\vec r})$ a spherical harmonics,
$\alpha_{1/2m_s}(\vec\sigma)$ a spin-1/2 function in spin space
and $\alpha_{1/2m_t}(\vec\tau)$ a spin-1/2 function in isospin space.
The single-particle variables are $x_i\equiv{\vec
r_i}{\vec\sigma_i}{\vec\tau_i}$
(a set of the corresponding radius-vector, spin and isospin variables).
The $e, l$ and $j$ are the principal, orbital and total angular momentum
quantum numbers, $m_l, m_s$ and $m_j$ are the magnetic projection quantum
numbers of orbital, spin and total angular momentum respectively.
The $m_t$ is projection of isospin defined such that $m_t=+1/2$
corresponds to a neutron state and $m_t=-1/2$ to a proton state.
A vector coupling of the angle and spin functions to form a state of good
total angular momentum is denoted by $\{\ldots\otimes\ldots\}_{jm_j}$.

Intrinsic properties of a nucleus can be described only by intrinsic wave
function. It depends on $(A-1)$ intrinsic (Jacobi) coordinates and
spin-isospin variables. In general eigenfunctions of Hamiltonian~(\ref{H_r})
could be represented as linear combination of products of intrinsic
wave functions and c.m. functions when all c.m. excitations are taken into
account.
Let us determine the system of orthonormalized Jacobi variables \cite{7,8}
according to the $2A-1$ vertices Jacobi tree presented in Fig. 1.
The upper vertices correspond to the single-particle variables
$x_1\ldots x_A$.
The remaining $A-1$ vertices (situated below the first ones) correspond to
the Jacobi variables which could be of three types:

\begin{itemize}
\item firstly if vertex associated with Jacobi variable is not connected
directly with any single--particle variable, the Jacobi variable is taken
to be equal the corresponding Jacobi coordinate
(as associated with $\xi_1\equiv\vec\xi_1$)
\item secondly that connected with only one of single--particle vertices
is associated with Jacobi variable which is composed of Jacobi coordinate
and directly reached single--particle spin-isospin variables
(as $\xi_3\equiv{\vec\xi_3}{\vec\sigma_{A-2}}{\vec\tau_{A-2}}$)
\item finally to that connected with two of single-particle vertices
will be prescribed Jacobi variable
which is composed of Jacobi coordinate and two sets of spin-isospin
variables coming from directly connected single-particle vertices
(as $\xi_2\equiv{\vec\xi_2}{\vec\sigma_{A-1}}{\vec\tau_{A-1}}
{\vec\sigma_A}{\vec\tau_A}$).
\end{itemize}

Orthogonal transformation to Jacobi coordinates is

\begin{eqnarray}
  \label{Jacobip}
  \cases{
    \vec \xi_\alpha = \displaystyle \sqrt{\frac{p_\alpha q_\alpha}{p_\alpha+q_\alpha}}
         \left[
           {1\over p_\alpha} \sum\limits_{j \in \{p_\alpha\}} \vec r_j -
           {1\over q_\alpha} \sum\limits_{j \in \{q_\alpha\}} \vec r_j
         \right]
        , & $\alpha = 1 \ldots A-1$  \cr
    \vec \xi_0 = \displaystyle \frac{1}{\sqrt{A}} \sum\limits_{j=1}^A \vec r_j
  }
\end{eqnarray}

Here $p_\alpha$ is the number of single-particle vertices which could be reached
while moving from the $\alpha$-th vertex upwards along the left edge;
$\{ p_\alpha\}$ is its numbers manifold and $q_\alpha,\{q_\alpha\}$--the
same for the right edge.
The $\vec\xi_0$ is proportional to the nucleus c.m. coordinate.
For the vertex connected with two last single-particle vertices we obtain for
instance $\vec \xi_2 = {1\over \sqrt{2}} (\vec r_{A-1} - \vec r_A)$.

\bigskip

\begin{picture}(300,220)
  \put(0,0){\line(1,0){300}}
   \put(0,220){\line(1,0){300}}
   \put(0,0){\line(0,1){220}}
   \put(300,0){\line(0,1){220}}
  \put(200,190){\line(2,-3){25}}
   \put(235,150){$\xi_2$}
  \put(90,130){\line(2,3){40}}
   \put(100,130){$\xi_4$}
  \put(104,109){\line(2,3){54}}
   \put(114,109){$\xi_3$}
  \put(50,190){\line(2,-3){100}}
   \put(48,195){$x_1$}
   \put(123,195){$x_{A-3}$}
   \put(153,195){$x_{A-2}$}
   \put(195,195){$x_{A-1}$}
   \put(245,195){$x_A$}
  \put(150,40){\line(2,3){100}}
   \put(160,40){$\xi_1$}
  \put(150,40){\line(0,-1){20}}
   \put(140,20){\llap{$\vec\xi_0$}}
  \multiput(80,195)(8,0){3}{.}
\end{picture}

\bigskip

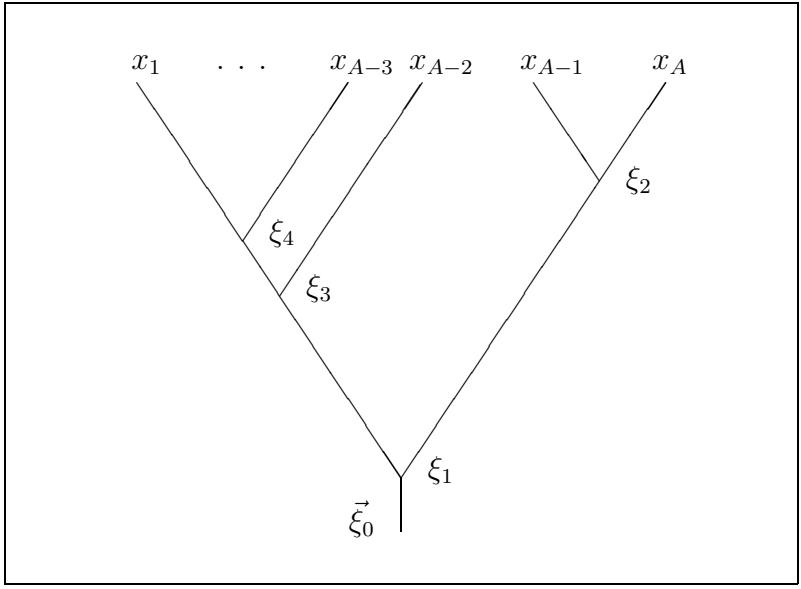
Fig.1 Jacobi tree definition

\bigskip

In the Jacobi coordinates the Hamiltonian~(\ref{H_r}) takes the form

\begin{eqnarray}
  \label{HoscJakobi}
  \begin{array}{r}
  \displaystyle H = \sum_{\alpha = 0}^{A-1}
       \left\{
        - \frac{\hbar^2}{2m}\Delta_{(\alpha)}
        + {1\over 2}m\omega^2
        \vec\xi_{\alpha}{\mathstrut}^2
       \right\}
       \hphantom{12345678901234567890123}
       \\
    \displaystyle =
       \left\{
        - \frac{\hbar^2}{2m}\Delta_{(0)}
        + {1\over 2}m\omega^2
        \vec\xi_{0}{\mathstrut}^2
       \right\}
       +
      \sum_{\alpha = 1}^{A-1}
       \left\{
        - \frac{\hbar^2}{2m}\Delta_{(\alpha)}
        + {1\over 2}m\omega^2
        \vec\xi_{\alpha}{\mathstrut}^2
       \right\}
       \\
    \displaystyle =
      h(\vec\xi_{0}) +
      \sum_{\alpha = 1}^{A-1} h(\vec\xi_{\alpha})
    \equiv
      H_{\mbox{\small c.m.\vphantom{intr}}} + H_{\mbox{\small intr.}}
  \end{array}
\end{eqnarray}

It implies that the Hamiltonian could be divided into two parts:
c.m. Hamiltonian $H_{\mbox{\small c.m.\vphantom{intr}}}$ and
Hamiltonian $H_{\mbox{\small intr.}}$ representing the intrinsic motion
of the nucleons in the system.
The set of $h(\vec\xi_{0})$'s eigenfunctions $\Psi_{elm}(\vec\xi_0)$ consist of
products of corresponding radial functions and spherical harmonics. Whereas
$h(\vec\xi_\alpha)$'s single-particle eigenfunctions could be of three
different types depending on number of single-particle variables directly
connected with $\alpha$-th vertex (if any):
\begin{itemize}
\item firstly there are functions depending on Jacobi coordinates which
have no direct connection with any single-particle coordinate (as $\vec
\xi_1$). Such the functions are products of radial function and
spherical harmonic
\item secondly there are functions of the form~(\ref{singl_f_r})
for Jacobi coordinates directly connected with one single-particle
coordinate (e.g. $\vec\xi_3$).
\item thirdly there will be functions depending on a Jacobi coordinate and
two sets of spin-isospin variables (e.g. $\xi_2$). They can be
represented as product of the vector coupled orbital and the vector coupled
two last particles in the list spin functions and vector coupled isospin
functions of corresponding particles.
\end{itemize}

 \begin{eqnarray}
     \psi_{elj\pi tm_jm_t}(\xi_2)
     =
     \left\{
       \psi_{el}(\vec\xi_2)
       \otimes
       \left\{
         \alpha_{1/2}(\vec\sigma_{A-1})
         \otimes
         \alpha_{1/2}(\vec\sigma_A)
       \right\}_s
     \right\}_{jm_j}
     \left\{
       \alpha_{1/2}(\vec\tau_{A-1})
       \otimes
       \alpha_{1/2}(\vec\tau_A)
     \right\}_{tm_t}
 \end{eqnarray}

Here orbital function $\psi_{elm}(\vec\xi_2)$ is product of the radial
function and spherical harmonic.
The intrinsic wavefunctions $\Psi_{E{\mit\Gamma}J{\mit\Pi}TM_JM_T}(\xi_1\ldots\xi_{A-1})$
are eigenfunctions of $H_{\mbox{\small intr.}}$. Since the antisymmetrization
procedure is rather cumbersome in Jacobi coordinates it is convenient to
introduce the expansion of the product of c.m. ground state function and
intrinsic wave-function in terms of shell model functions

 \begin{eqnarray}
   \label{trpersl}
   \Psi_{00}(\vec\xi_0)
   \Psi_{E{\mit\Gamma}J{\mit\Pi}TM_JM_T}(\xi_1\ldots\xi_{A-1})
   =
   \sum_{K{\mit\Delta}}
   \Psi_{EK{\mit\Delta}J{\mit\Pi}TM_JM_T}(x_1\ldots x_A)
   \; a_{K{\mit\Delta};00,{\mit\Gamma}}^{EJ{\mit\Pi}T}
 \end{eqnarray}

The coefficients of this expansion
$a_{K{\mit\Delta};00,{\mit\Gamma}}^{EJ{\mit\Pi}T}$ can be obtained
diagonalizing c.m. Hamiltonian $H_{\mbox{\small c.m.\vphantom{intr}}}$
in the shell model basis.
The summation in this formula runs over all configurations $K$
and additional quantum number $\mit\Delta$ (in spirit of the ${\mit\Gamma}$).
Here zeros indicate the c.m. ground state: principal and orbital angular
momentum quantum numbers.
The intrinsic wave-function depends on
the $5A - 3$ intrinsic Jacobi variables $\xi_1\ldots\xi_{A-1}$ where
$\xi_\alpha$ stands for the $\vec\xi_\alpha$ together with one of the
mentioned three types of the sets of spin and isospin variables.

Due to the identity of the nucleons it is convenient to restrict consideration
only by two last particles in the list.
General form of an expression of two-particle intrinsic density matrix is
chosen to be

 \begin{eqnarray}
   \label{aQJacobi}
   \begin{array}{r}
     Q^{E{\mit\Gamma}J{\mit\Pi}T,E'{\mit\Gamma}'J{\mit\Pi}T}(\xi_2,\xi_2')
     =\displaystyle{1\over[J,T]}
     \sum\limits_{M_J,M_T}
     \int d\vec\xi_0d\xi_1d\xi_3\ldots d\xi_{A-1}
     \Psi_{00}(\vec\xi_0)\Psi_{00}^*(\vec\xi_0)
     \\
     \times
     \Psi_{E{\mit\Gamma}J{\mit\Pi}TM_JM_T}(\xi_1,\xi_2\ldots\xi_{A-1})
     \Psi_{E'{\mit\Gamma}'J{\mit\Pi}TM_JM_T}^*(\xi_1,\xi_2'\ldots\xi_{A-1})
   \end{array}
 \end{eqnarray}

where integration sign denotes integration over the continuous variables of
the relative motion of $A-2$ particles and summation over discrete variables
of the corresponding nucleons according to the Jacobi tree in use.
Here and in the following the notations as $[J,T]\equiv (2J+1)(2T+1)$ is
shortcut of number of states with corresponding angular momentum and isospin.
The most convenient way to construct an antisymmetric wave function is the
method of fractional parentage coefficients (CFP) developed in \cite{4}.

 \begin{eqnarray}
   \label{trperTISMCFP}
   \begin{array}{r}
     \Psi_{E{\mit\Gamma}J{\mit\Pi}TM_JM_T}(\xi_1\ldots\xi_{A-1})
     = \displaystyle
     \sum\limits_{\stackrel{\scriptstyle\overline{\overline{(E{\mit\Gamma}J{\mit\Pi}T)}}}
                    {\scriptstyle elj\pi t}
     }
     [\overline{\overline{(E{\mit\Gamma}J{\mit\Pi}T)}};
     elj\pi t|| E{\mit\Gamma}J{\mit\Pi}T]
     \\
     \times
     \left\{
       \Psi_{\overline{\overline{E{\mit\Gamma}J{\mit\Pi}T}}}(\xi_1\ldots\xi_{A-1})
       \otimes
       \psi_{elj\pi t}(\xi_2)
     \right\}_{J{\mit\Pi}TM_JM_T}
   \end{array}
 \end{eqnarray}

Here double bar indicates the grandparent state and $elj\pi t$ characterize
the separated subsystem.
The summation in this formula spans subspaces of grandparent
$\overline{\overline{(E{\mit\Gamma}J{\mit\Pi}T)}}$ and two particle
$elj\pi t$ states which satisfy the necessary selection rules and the energy,
momentum and parity conservation conditions.
The expansion coefficients have to be intrinsic CFP. By means
of~(\ref{trperTISMCFP}) it is possible to get the expression
of the intrinsic density matrix

 \begin{eqnarray}
   \label{QperTISMCFP}
   \begin{array}{r}
     Q^{E{\mit\Gamma}J{\mit\Pi}T,E'{\mit\Gamma}'J{\mit\Pi}T}(\xi_2,\xi_2')
     =
     \displaystyle\sum\limits_{ele'l'j\pi t}
     \sum\limits_{\overline{\overline{(E{\mit\Gamma}J{\mit\Pi}T)}}}
     [\overline{\overline{(E{\mit\Gamma}J{\mit\Pi}T)}};
     elj\pi t|| E{\mit\Gamma}J{\mit\Pi}T]
     \\
     \times\displaystyle
     [\overline{\overline{(E{\mit\Gamma}J{\mit\Pi}T)}};
     e'l'j\pi t|| E'{\mit\Gamma}'J{\mit\Pi}T]
     \;
     {1\over[j,t]}
     \sum\limits_{m_j,m_t}
     \psi_{elj\pi tm_jm_t}(\xi_2)
     \psi_{e'l'j\pi tm_jm_t}^*(\xi_2')
   \end{array}
 \end{eqnarray}

The intrinsic density matrix in harmonic oscillator representation is

 \begin{eqnarray}
   \begin{array}{r}
     W_{elj\pi t,e'l'j\pi t}^{E{\mit\Gamma}J{\mit\Pi}T,E'{\mit\Gamma}'J{\mit\Pi}T}
     =
     \displaystyle
     \sum\limits_{\overline{\overline{(E{\mit\Gamma}J{\mit\Pi}T)}}}
     [\overline{\overline{(E{\mit\Gamma}J{\mit\Pi}T)}};
     elj\pi t|| E{\mit\Gamma}J{\mit\Pi}T]
     [\overline{\overline{(E{\mit\Gamma}J{\mit\Pi}T)}};
     e'l'j\pi t|| E'{\mit\Gamma}'J{\mit\Pi}T]
   \end{array}
 \end{eqnarray}

This matrix could be defined avoiding the appearance of intrinsic CFP.
The key is relation~(\ref{trpersl})
between intrinsic and shell model wave functions.
The usual shell model fractional parentage expansion could be used to separate
out two last particles in the list

 \begin{eqnarray}
   \begin{array}{r}
     \Psi_{EK{\mit\Delta}J{\mit\Pi}TM_JM_T}(x_1\ldots x_A)
     = \hphantom{123456789012345678901234567890123456789012345}
     \\
     =
     \displaystyle
     \sum\limits_{\stackrel{\scriptstyle\overline{\overline{(EK{\mit\Delta}J{\mit\Pi}T)}}}
                    {\scriptstyle (elj)_{A-1},(elj)_A,J''T''}
     }
     \langle \overline{\overline{(EK{\mit\Delta}J{\mit\Pi}T)}};
     ((elj)_{A-1},(elj)_A)J''T''|| EK{\mit\Delta}J{\mit\Pi}T\rangle
     \\
     \times
     \left\{
       \Psi_{\overline{\overline{(EK{\mit\Delta}J{\mit\Pi}T)}}}\, (x_1\ldots x_{A-2})
       \otimes
       \Phi_{((elj)_{A-1},(elj)_A)J''T''}(x_{A-1},x_A)
     \right\}_{J{\mit\Pi}TM_JM_T}
   \end{array}
 \end{eqnarray}

Here inverted commas refer to the separated two-particle subsystem.
For the sake of convenience of the notations we shall use comma to denote
the nucleons variables with respect to which a wave-function is antisymmetric.
The coefficients of introduced expansion (generalized CFP (GCFP)
as defined by Levinson \cite{9}) enable to express the antisymmetric
shell model wave function of $A$ nucleons in the form of linear combinations
of products of antisymmetric functions of $A - 2$ nucleons and antisymmetric
two-particle wave functions.
The separation of the two nucleons from the initial configuration
can be accomplished in all possible ways consistent with required
triangular relations, thus
giving rise to transformation matrix describing transformation between
different momentum coupling schemes.
In case when two nucleons are taken from different shells the GCFP
can be expressed in terms of single shell one-particle  CFP and
corresponding transformation matrix

 \begin{eqnarray}
   \begin{array}{r}
     \langle \overline{\overline{(EK{\mit\Delta}J{\mit\Pi}T)}};
     ((elj)_{A-1},(elj)_A)J''T''|| EK{\mit\Delta}J{\mit\Pi}T\rangle
     =
     \displaystyle(-1)^{\nu_{\scriptstyle r}+\nu_{\scriptstyle p}-1}
     \left(\frac{2n_rn_p}{A(A-1)}\right)^{1\over2}
     \hphantom{1234}
     \\
     \times
     \langle(elj)_r^{n_{\scriptstyle r}-1}\overline{({\mit\Delta}JT)}_r;
     (elj)_r||(elj)_r^{n_{\scriptstyle r}}({\mit\Delta}JT)_r\rangle
     \langle(elj)_p^{n_{\scriptstyle p}-1}\overline{({\mit\Delta}JT)}_p;
     (elj)_p||(elj)_p^{n_{\scriptstyle p}}({\mit\Delta}JT)_p\rangle
     \\
     \times
     \langle
       ((J_1T_1\ldots\overline{J_rT_r}\ldots\overline{J_pT_p}\ldots
       J_kT_k)\overline{\overline{JT}}, (j_rt_r,j_pt_p)J''T'')JT|
       \hphantom{1234567890123456789012}
       \\
       |(J_1T_1\ldots(\overline{J_rT_r},j_rt_r)J_rT_r\ldots
       (\overline{J_pT_p},j_pt_p)J_pT_p\ldots J_kT_k)^K JT
     \rangle
   \end{array}
 \end{eqnarray}

Here single bar over the quantum numbers indicates the parent state,
subscript $r$ refers to the $r$-th shell in the configuration and
superscript $n_r$ is number of particles contained in the $r$-th shell.
The integer number $\displaystyle\nu_{\scriptstyle r} = \sum_{i = r+1}^k n_i$,
where sum runs over all shells standing to the right
from the $r$-th shell.
The $\langle(elj)_r^{n_{\scriptstyle r}-1}\overline{({\mit\Delta}JT)}_r;
(elj)_r||(elj)_r^{n_{\scriptstyle r}}({\mit\Delta}JT)_r\rangle$
denotes the one-particle CFP of $r$-th shell.
When two nucleons are separated from the same shell the
GCFP definition contains two-particle CFP:

 \begin{eqnarray}
   \begin{array}{r}
     \langle \overline{\overline{(EK{\mit\Delta}J{\mit\Pi}T)}};
     ((elj)_{A-1},(elj)_A)J''T''|| EK{\mit\Delta}J{\mit\Pi}T\rangle
     =\hphantom{1234567890123456789012345678}
     \\
     =
     \displaystyle
     \left(\frac{n_r(n_r-1)}{A(A-1)}\right)^{1\over2}
     \langle(elj)_r^{n_{\scriptstyle r}-2}\overline{\overline{({\mit\Delta}JT)}}_r;
     (elj)_r^2J''T''||(elj)_r^{n_{\scriptstyle r}}({\mit\Delta}JT)_r\rangle
     \\
     \times
     \langle
       ((J_1T_1\ldots\overline{\overline{J_rT_r}}\ldots
       J_kT_k)\overline{\overline{JT}}, J''T'')JT|
        (J_1T_1\ldots(\overline{\overline{J_rT_r}},J''T'')J_rT_r\ldots J_kT_k)^K JT
     \rangle
   \end{array}
 \end{eqnarray}

The simplest way of CFP matrix calculation is proposed in \cite{4}.
The method is based on the observation that the spectral decomposition
of antisymmetrization operator matrix is not uniquely defined.
The best choice is to set the upper triangle of CFP matrix equal to zero.
This method was implemented
in the program code and all s-d shells CFP were calculated. Corresponding
two-particle CFP were obtained following the well-known Redmond formula.
Transformation matrix describing momentum recoupling was calculated using
its direct representation by sum of Clebsh-Gordan coefficients.

The intrinsic Jacobi variable $\xi_2$ could be introduced by expressing
antisymmetric two-particle wave-functions
$\Phi_{((elj)_{A-1},(elj)_A)J''T''}(x_{A-1},x_A)$
in terms of single-particle eigenfunctions~(\ref{singl_f_r}) and coupling
them in proper order for well-known Talmi-Moshinsky transformation to apply.
At first the antisymmetric wave functions
$\Phi_{((elj)_{A-1},(elj)_A)J''T''}(x_{A-1},x_A)$
should be expressed as a linear combination of not antisymmetrized coupled
momentum wave-functions.
Here and below we will
indicate such functions with semicolon as separation mark for the variables
in the list. In case when the nucleons are taken from the same shell we have

 \begin{eqnarray}
   \begin{array}{r}
     \Phi_{((elj)_{A-1},(elj)_A)J''T''}(x_{A-1},x_A)
     =
     \displaystyle
     {1\over2}\left[1-(-1)^{J''+T''}\right]
     \Phi_{((elj)_{A-1};(elj)_A)J''T''}(x_{A-1};x_A)
   \end{array}
 \end{eqnarray}

Here we suppress for brevity magnetic quantum numbers.
When the two nucleons are taken from different shells linear combination
is of the form

 \begin{eqnarray}
   \begin{array}{r}
     \Phi_{((elj)_{A-1},(elj)_A)J''T''}(x_{A-1},x_A)
     =
     \displaystyle
     {1\over\sqrt{2}}
     \left[\Phi_{((elj)_{A-1};(elj)_A)J''T''}(x_{A-1};x_A)\right.
     \\
     \left. -(-1)^{j_{A-1}+j_A+1-J''-T''}
     \Phi_{((elj)_A;(elj)_{A-1})J''T''}(x_{A-1};x_A)\right]
   \end{array}
 \end{eqnarray}

We reveal the complete representation of the coupled momentum wave functions
by the coupled orbital-spin and isospin functions

 \begin{eqnarray}
   \begin{array}{r}
     \Phi_{((elj)_{A-1};(elj)_A)J''T''M_J''M_T''}(x_{A-1};x_A)
     =
     \left\{
       \alpha_{1/2}(\vec\tau_{A-1})
       \otimes
       \alpha_{1/2}(\vec\tau_A)
     \right\}_{T''M_T''}
     \hphantom{123456789012345}
     \\
     \times
     \left\{
       \left\{
         \phi_{e_{A-1}l_{A-1}}(\vec r_{A-1})
         \otimes
         \alpha_{1/2}(\vec\sigma_{A-1})
       \right\}_{j_{A-1}}
       \otimes
       \left\{
         \phi_{e_Al_A}(\vec r_A)
         \otimes
         \alpha_{1/2}(\vec\sigma_A)
       \right\}_{j_A}
     \right\}_{J''M_J''}
   \end{array}
 \end{eqnarray}

Here orbital function is defined as
$\phi_{elm}(\vec r)=R_{el}(r)Y_{lm}(\widehat{\vec r})$.

Second point concerns transition to the appropriate Jacobi coordinates.
This transformation from single-particle variables $x_{A-1}$ and $x_A$
to Jacobi variable $\xi_2$ could be accomplished by assistance
of Jacobi coordinates with nonpositive indices \cite{8}. The laters are
chosen to be proportional to the c.m. of corresponding subsystems.

\begin{eqnarray}
    \vec \xi_{1-\alpha} =
      \displaystyle\frac{1}{\sqrt{p_\alpha+q_\alpha}}
      \sum\limits_{j \in \{p_\alpha\bigcap q_\alpha\}} \vec r_j
        , \alpha = 1 \ldots A-1
\end{eqnarray}

For example $\vec\xi_2$ is accompanied with
$\vec \xi_{-1} = {1\over \sqrt{2}} (\vec r_{A-1} + \vec r_A)$
which is simply proportional to the c.m. of the two last particles.
When $\alpha=1$ corresponding Jacobi coordinate with nonpositive index
coincide with $\vec \xi_0$ defined in~(\ref{Jacobip}).
This orthogonal transformation allowing to climb up and down the
Jacobi tree is defined

\begin{eqnarray}
  \label{Jacobim}
  \left\{
    \begin{array}{lll}
    \vec \xi_\alpha &=& \displaystyle \sqrt{1\over{1+d_\alpha}}\;
    \vec\xi_{-\mu}-
                   \sqrt{d_\alpha\over{1+d_\alpha}}\; \vec \xi_{-\nu}
    \vphantom{\Biggl(} \\
    \vec \xi_{1-\alpha} &=& \displaystyle \sqrt{d_\alpha\over{1+d_\alpha}}\;
    \vec\xi_{-\mu}+
                       \sqrt{1\over{1+d_\alpha}}\; \vec \xi_{-\nu}
    \vphantom{\Biggl(}
    \end{array}
  \right.
\end{eqnarray}

Here $d_\alpha=p_{\alpha}/q_{\alpha}$, $\vec\xi_{-\mu}$ is Jacobi
coordinate associated with vertex which could be reached while moving from
the $\alpha$-th vertex upwards along the left edge;
$\vec \xi_{-\nu}$ is the same for the right edge.
In case when any of the upper vertices is bound to single-particle variables
the corresponding single-particle coordinate is taken.

To apply this transformation it is necessary to interchange the order of
coupling. It will be the so-called L-S coupling scheme.

 \begin{eqnarray}
   \label{sldvbet}
   \begin{array}{r}
     \left\{
       \left\{
         \phi_{e_{A-1}l_{A-1}}(\vec r_{A-1})
         \otimes
         \alpha_{1/2}(\vec\sigma_{A-1})
       \right\}_{j_{A-1}}
       \otimes
       \left\{
         \phi_{e_Al_A}(\vec r_A)
         \otimes
         \alpha_{1/2}(\vec\sigma_A)
       \right\}_{j_A}
     \right\}_{J''M_J''}
     =
     \\
     =
     \displaystyle\sum\limits_{Ls}
     \langle((l_{A-1},{\scriptstyle 1/2})j_{A-1},(l_A,{\scriptstyle 1/2})j_A)J''|
     ((l_{A-1},l_A)L,({\scriptstyle 1/2,1/2})s)J''\rangle
     \\
     \times
     \left\{
       \left\{
         \phi_{e_{A-1}l_{A-1}}(\vec r_{A-1})
         \otimes
         \phi_{e_Al_A}(\vec r_A)
       \right\}_L
       \otimes
       \left\{
         \alpha_{1/2}(\vec\sigma_{A-1})
         \otimes
         \alpha_{1/2}(\vec\sigma_A)
       \right\}_s
     \right\}_{J''M_J''}
   \end{array}
 \end{eqnarray}

Here $L$ is the orbital momentum of the relative motion of the two
last nucleons and $s$ is the corresponding spin.
Now can be used the Talmi - Moshinsky - Smirnov transformation

 \begin{eqnarray}
   \label{Talmi}
   \begin{array}{r}
     \left\{
       \phi_{e_{A-1}l_{A-1}}(\vec r_{A-1})
       \otimes
       \phi_{e_Al_A}(\vec r_A)
     \right\}_{LM}
     = \hphantom{123456789012345678901234567890}
     \\
     =
     \displaystyle\sum\limits_{(el)_{-1},el}
     \langle(el)_{A-1},(el)_A:L|(el)_{-1},el:L\rangle_1
     \left\{
       \psi_{(el)_{-1}}(\vec \xi_{-1})
       \otimes
       \psi_{el}(\vec\xi_2)
     \right\}_{LM}
   \end{array}
 \end{eqnarray}

here sum is restricted under energy $(e_{A-1} + e_A = e_{-1} + e)$ and parity
conservation.
The Talmi - Moshinsky - Smirnov coefficients
$\langle(el)_{A-1},(el)_A:L|(el)_{-1},el:L\rangle_{d_\alpha}$
following \cite{10,11}
are the elements of the matrix
for the transition between the oscillator functions depending on coordinates
which are related with orthogonal transformation~(\ref{Jacobim})~.
Finally we must return to the momentum coupling defined in
the $\psi_{elj\pi tm_jm_t}(\xi_2)$ functions

 \begin{eqnarray}
   \label{trdvbet}
   \begin{array}{r}
     \left\{
       \left\{
         \psi_{(el)_{-1}}(\vec \xi_{-1})
         \otimes
         \psi_{el}(\vec\xi_2)
       \right\}_L
       \otimes
       \left\{
         \alpha_{1/2}(\vec\sigma_{A-1})
         \otimes
         \alpha_{1/2}(\vec\sigma_A)
       \right\}_s
     \right\}_{J''M_J''}
     = 
     \\
     =
     \displaystyle\sum\limits_j
     \langle((l_{-1},l)L,s)J''|(l_{-1},(l,s)j)J''\rangle
     \\
     \times
     \left\{
       \psi_{(el)_{-1}}(\vec \xi_{-1})
       \otimes
       \left\{
         \psi_{el}(\vec\xi_2)
         \otimes
         \left\{
           \alpha_{1/2}(\vec\sigma_{A-1})
           \otimes
           \alpha_{1/2}(\vec\sigma_A)
         \right\}_s
       \right\}_j
     \right\}_{J''M_J''}
   \end{array}
 \end{eqnarray}

Now we can express functions from the left side of the
expression~(\ref{sldvbet}) in terms of the functions from the right side
of the expression~(\ref{trdvbet})

 \begin{eqnarray}
   \begin{array}{r}
     \left\{
       \left\{
         \phi_{e_{A-1}l_{A-1}}(\vec r_{A-1})
         \otimes
         \alpha_{1/2}(\vec\sigma_{A-1})
       \right\}_{j_{A-1}}
       \otimes
       \left\{
         \phi_{e_Al_A}(\vec r_A)
         \otimes
         \alpha_{1/2}(\vec\sigma_A)
       \right\}_{j_A}
     \right\}_{J''M_J''}
     =
     \\
     =
     \displaystyle\sum\limits_{(el)_{-1},elsj}
     \langle((elj)_{A-1};(elj)_A)J''|((el)_{-1},elsj)J''\rangle
     \\
     \times
     \left\{
       \psi_{(el)_{-1}}(\vec \xi_{-1})
       \otimes
       \left\{
         \psi_{el}(\vec\xi_2)
         \otimes
         \left\{
           \alpha_{1/2}(\vec\sigma_{A-1})
           \otimes
           \alpha_{1/2}(\vec\sigma_A)
         \right\}_s
       \right\}_j
     \right\}_{J''M_J''}
   \end{array}
 \end{eqnarray}

The coefficients of introduced expansion represents the collection of the
transformation matrices from~(\ref{trdvbet}) and (\ref{sldvbet}),
and the Talmi - Moshinsky coefficients from the expansion~(\ref{Talmi})

 \begin{eqnarray}
   \begin{array}{r}
     \langle((elj)_{A-1};(elj)_A)J''|((el)_{-1},elsj)J''\rangle
     =
     \displaystyle\sum\limits_{L}
     \langle((l_{-1},l)L,s)J''|(l_{-1},(l,s)j)J''\rangle
     \\
     \times
     \langle((l_{A-1},{\scriptstyle 1/2})j_{A-1},(l_A,{\scriptstyle 1/2})j_A)J''|
     ((l_{A-1},l_A)L,({\scriptstyle 1/2,1/2})s)J''\rangle
     \\
     \times
     \langle(el)_{A-1},(el)_A:L|(el)_{-1},el:L\rangle_1
   \end{array}
 \end{eqnarray}

Corresponding transformation matrices can be represented
in terms of standard vector coefficients: 6-j and 9-j, thus we are led to
the expression

\begin{eqnarray}
  \begin{array}{r}
    \langle((elj)_{A-1};(elj)_A)J''|((el)_{-1},elsj)J''\rangle =
    (-1)^{l_{A-1}+l_A+s+J''}\sqrt{\left[ j_{A-1},j_A,s,j\right]}
    \hphantom{1234567890}
    \\
    \times
    \displaystyle\sum\limits_L [L]
    \langle((el)_{A-1},(el)_A):L|(el)_{-1},el:L\rangle_1
    \left\{
      \matrix{
        l_{-1} & l   & L \cr
        s      & J'' & j \cr
      }
    \right\}
    \left\{
      \matrix{
        l_{A-1} & {1\over2} & j_{A-1} \cr
        l_A     & {1\over2} & j_A \cr
        L       & s         & J'' \cr
      }
    \right\}
  \end{array}
\end{eqnarray}

It should be noted that presented coefficients weighted only not
antisymmetrical functions as is stressed by semicolon.
Taking into account the isospin part of the two-particle function
we get the final form of the coefficients for the transition from
the antisymmetrical two-particle shell model functions to the sought
$\psi_{elj\pi tm_jm_t}(\xi_2)$ vector coupled with the
introduced functions $\psi_{(el)_{-1}}(\vec \xi_{-1})$

\begin{eqnarray}
  \label{Jacobi2}
  \begin{array}{r}
    \langle((elj)_{A-1},(elj)_A)J''T''|((el)_{-1},elj\pi t)J''T''\rangle
    =
    \displaystyle\frac{\displaystyle\delta_{t,T''}\left[1-(-1)^{l+s+t}\right]}
    {\displaystyle\sqrt{2\left(1+\delta_{(\varepsilon lj)_{N-1},(\varepsilon lj)_N)}\right)}}
    \\
    \times
    \langle((elj)_{A-1};(elj)_A)J''|((el)_{-1},elsj)J''\rangle
  \end{array}
\end{eqnarray}

This transformation allows us to re-express the two-particle shell model
function expansion in the form

 \begin{eqnarray}
   \begin{array}{r}
     \Phi_{((elj)_{A-1},(elj)_A)J''T''M_J''M_T''}(x_{A-1},x_A)
     =
     \displaystyle\sum\limits_{(el)_{-1},elsj}
     \left\{
       \alpha_{1/2}(\vec\tau_{A-1})
       \otimes
       \alpha_{1/2}(\vec\tau_A)
     \right\}_{T''M_T''}
     \\
     \times
     \left\{
       \psi_{(el)_{-1}}(\vec \xi_{-1})
       \otimes
       \left\{
         \psi_{el}(\vec\xi_2)
         \otimes
         \left\{
           \alpha_{1/2}(\vec\sigma_{A-1})
           \otimes
           \alpha_{1/2}(\vec\sigma_A)
         \right\}_s
       \right\}_j
     \right\}_{J''M_J''}
     \\
     \times
     \langle((elj)_{A-1},(elj)_A)J''T''|((el)_{-1},elj\pi t)J''T''\rangle
   \end{array}
 \end{eqnarray}

Finally we obtain the eigenfunctions of $A$--particles
Hamiltonian~(\ref{H_r})
with separated two groups of nucleons: first, containing
two last particles which are described by antisymmetrical, coupled function,
depending on Jacobi coordinates, and second, containing remaining set of
particles, which are described by corresponding eigenfunction of
$A-2$--particles Hamiltonian depending on single-particle coordinates.
That functions we call the
A--particle oscillator functions with singled out dependence on intrinsic
coordinates of two last particles in the list.

 \begin{eqnarray}
   \label{slperm}
   \begin{array}{r}
     \Psi_{EK{\mit\Delta}J{\mit\Pi}TM_JM_T}(x_1\ldots x_A)
     = \hphantom{123456789012345678901234567890123456789012345}
     \\
     =
     \displaystyle\sum\limits_{\overline{\overline{(EK{\mit\Delta}J{\mit\Pi}T)}}}
     \sum\limits_{(el)_{-1}} \sum\limits_{elj\pi t,J''T''}
     \langle\overline{\overline{(EK{\mit\Delta}J{\mit\Pi}T)}};
     ((el)_{-1},elj\pi t)J''T''|| EK{\mit\Delta}J{\mit\Pi}T\rangle
     \\
     \times
     \left\{
       \Psi_{\overline{\overline{(EK{\mit\Delta}J{\mit\Pi}T)}}}\, (x_1\ldots x_{A-2})
       \otimes
       \left\{
         \psi_{(el)_{-1}}(\vec \xi_{-1})
         \otimes
         \psi_{elj\pi t}(\xi_2)
       \right\}_{J''T''}
     \right\}_{J{\mit\Pi}TM_JM_T}
   \end{array}
 \end{eqnarray}

The momentum coupling scheme between the functions can be
arbitrary since density matrix is invariant with respect to orthogonal
their transformation. Taking advantage of this will
save us for one more recoupling looked for coupling scheme with
$\psi_{elj\pi t}(\xi_2)$ coupled at end.

The coefficients of the~(\ref{slperm}) functions expansion by introduced
ones are composed of GCFP and coefficients~(\ref{Jacobi2})

 \begin{eqnarray}
   \begin{array}{r}
     \langle \overline{\overline{(EK{\mit\Delta}J{\mit\Pi}T)}};
     ((el)_{-1},elj\pi t)J''T''|| EK{\mit\Delta}J{\mit\Pi}T\rangle
     = \hphantom{123456789012345678901234567890}
     \\
     =
     \displaystyle\sum\limits_{(elj)_{A-1},(elj)_A}
     \langle \overline{\overline{(EK{\mit\Delta}J{\mit\Pi}T)}};
     ((elj)_{A-1},(elj)_A)J''T''|| EK{\mit\Delta}J{\mit\Pi}T\rangle
     \\
     \times
     \langle((elj)_{A-1},(elj)_A)J''T''|((el)_{-1},elj\pi t)J''T''\rangle
   \end{array}
 \end{eqnarray}

This set of coefficients may satisfy the orthonormalization condition

 \begin{eqnarray}
   \begin{array}{r}
     \displaystyle\sum\limits_{\overline{\overline{(EK{\mit\Delta}J{\mit\Pi}T)}}}
     \sum\limits_{(el)_{-1}} \sum\limits_{elj\pi t,J''T''}
     \langle\overline{\overline{(EK{\mit\Delta}J{\mit\Pi}T)}};
     ((el)_{-1},elj\pi t)J''T''|| EK{\mit\Delta}J{\mit\Pi}T\rangle
     \\
     \times
     \langle\overline{\overline{(EK{\mit\Delta}J{\mit\Pi}T)}};
     ((el)_{-1},elj\pi t)J''T''|| EK'{\mit\Delta}'J{\mit\Pi}T\rangle
     =
     \delta_{K{\mit\Delta},K'{\mit\Delta}'}
   \end{array}
 \end{eqnarray}

Now we are able to get the final expression of intrinsic density matrix.
Preliminarily it should be noted that due to
orthogonality of transformations~(\ref{Jacobip}) and (\ref{Jacobim})
it may hold equivalency for integration over two sets of variables

 \begin{eqnarray}
   \label{intkint}
   \int d\vec\xi_0d\xi_1d\xi_3\ldots d\xi_{A-1}
   \doteq
   \int dx_1\ldots dx_{A-1}d\vec\xi_{-1}
 \end{eqnarray}

Here the integration is taken in sense of the definition~(\ref{aQJacobi}).
Inserting in the  density matrix~(\ref{aQJacobi})
the linear combination of shell model functions with fixed c.m. state
(\ref{trpersl}) and integrating after substitution of variables
according~(\ref{intkint}) we get the sought intrinsic density matrix
expression

 \begin{eqnarray}
   \label{Qpersl}
   \begin{array}{r}
     Q^{E{\mit\Gamma}J{\mit\Pi}T,E'{\mit\Gamma}'J{\mit\Pi}T}(\xi_2,\xi_2')
     =
     \displaystyle\sum\limits_{ele'l'j\pi t}
     \sum_{K{\mit\Delta},K'{\mit\Delta}'}
     a_{K{\mit\Delta};00,{\mit\Gamma}}^{EJ{\mit\Pi}T}
     a_{K'{\mit\Delta}';00,{\mit\Gamma}'}^{E'J{\mit\Pi}T}
     \\
     \times\displaystyle
     \sum\limits_{\stackrel{\scriptstyle\overline{\overline{(EK{\mit\Delta}J{\mit\Pi}T)}}}
                    {\scriptstyle (el)_{-1}J''T''}
     }
     \langle \overline{\overline{(EK{\mit\Delta}J{\mit\Pi}T)}};
     ((el)_{-1},elj\pi t)J''T''|| EK{\mit\Delta}J{\mit\Pi}T\rangle
     \\
     \times\displaystyle
     \langle \overline{\overline{(EK{\mit\Delta}J{\mit\Pi}T)}};
     ((el)_{-1},e'l'j\pi t)J''T''|| E'K'{\mit\Delta}'J{\mit\Pi}T\rangle
     \\
     \times\displaystyle
     {1\over[j,t]}
     \sum\limits_{m_j,m_t}
     \psi_{elj\pi tm_jm_t}(\xi_2)
     \psi_{elj\pi tm_jm_t}^*(\xi_2')
   \end{array}
 \end{eqnarray}

Comparing the expressions at the products of the functions
$\psi_{elj\pi tm_jm_t}(\xi_2)\psi_{elj\pi tm_jm_t}^*(\xi_2')$
in the intrinsic density matrix expressions (\ref{QperTISMCFP}) and
(\ref{Qpersl}) we are led to the final result

 \begin{eqnarray}
   \begin{array}{r}
     W_{elj\pi t,e'l'j\pi t}^{E{\mit\Gamma}J{\mit\Pi}T,E'{\mit\Gamma}'J{\mit\Pi}T}
     =
     \displaystyle
     \sum\limits_{\stackrel{\scriptstyle\overline{\overline{(EK{\mit\Delta}J{\mit\Pi}T)}}}
                    {\scriptstyle (el)_{-1}J''T''}
     }
     \langle \overline{\overline{(EK{\mit\Delta}J{\mit\Pi}T)}};
     ((el)_{-1},elj\pi t)J''T''|| E{\mit\Gamma}J{\mit\Pi}T\rangle
     \\
     \times
     \langle \overline{\overline{(EK{\mit\Delta}J{\mit\Pi}T)}};
     ((el)_{-1},e'l'j\pi t)J''T''|| E'{\mit\Gamma}'J{\mit\Pi}T\rangle
   \end{array}
 \end{eqnarray}

Presented coefficients are connected with outlined above coefficients
(\ref{slperm}) and (\ref{trpersl}) according to the relation

 \begin{eqnarray}
   \label{koefslitr}
   \begin{array}{r}
     \langle \overline{\overline{(EK{\mit\Delta}J{\mit\Pi}T)}};
     ((el)_{-1},elj\pi t)J''T''|| E{\mit\Gamma}J{\mit\Pi}T\rangle
     = \hphantom{123456789012345678901234567890}
     \\
     =
     \displaystyle\sum\limits_{K{\mit\Delta}}
     \langle \overline{\overline{(EK{\mit\Delta}J{\mit\Pi}T)}};
     ((el)_{-1},elj\pi t)J''T''|| EK{\mit\Delta}J{\mit\Pi}T\rangle
     \; a_{K{\mit\Delta};00,{\mit\Gamma}}^{EJ{\mit\Pi}T}
   \end{array}
 \end{eqnarray}

That coefficients project out the space of
A--particle oscillator functions with singled out dependence on intrinsic
coordinates of two last particles into
intrinsic functions subspace with not excited \mbox{c.m..}

The intrinsic density matrix in harmonic oscillator representation satisfy
a usual normalization relation

 \begin{eqnarray}
   \begin{array}{r}
     \displaystyle\sum\limits_{elj\pi t}
     W_{elj\pi t,elj\pi t}^{E{\mit\Gamma}J{\mit\Pi}T,E'{\mit\Gamma}'J{\mit\Pi}T}
     =
     \delta_{E{\mit\Gamma},E'{\mit\Gamma}'}
   \end{array}
 \end{eqnarray}

The presented expressions enable to obtain the full set of intrinsic
density matrices describing A--nucleon system in isospin formalism
after calculating the coefficients (\ref{koefslitr}) weighting linear
combination of that eigenfunctions of $A$--particles
Hamiltonian~(\ref{H_r}) for which ones c.m. is in fixed state.

\section{COMPUTATIONAL RESULTS}

Since momentum recoupling, antisymmetrization and transformation to Jacobi
coordinates are orthogonal transformations a precise arithmetics could be
applied. Instead of calculations with real numbers, which are connected
with serious numerical instabilities, calculations were performed with
numbers represented in the form $n/(m\sqrt{k})$, where n, m and k are
integers.

On the basis of precise arithmetics were developed computational
procedures of a number of well-known transformations coefficients and
implemented in a computer code. As follows:
one-particle single shell CFP \cite{5},
two-particle single shell CFP according to Redmond expression,
Clebsh-Gordan coefficients,
6-j and 9-j vector coupling coefficients,
momentum recoupling matrices,
GCFP,
Talmi coefficients.

Another set of computational procedures was developed for enumeration
of the antisymmetric states by means of combinatorial calculations.

Outlined above computational procedures were used for general formalism,
presented in the section Number 2, to implement in a computer code.
For illustration let us propose the calculations of 6, 7 and 8 nucleon
systems with minimal oscillator
energy compatible with the Pauli exclusion principle were performed.
The number of calculated intrinsic CFP and intrinsic density matrices
for 6, 7 and 8 nucleon systems are: 255 and 41,
1345 and 66, 5021 and 138 respectively.

As an example we consider the case A = 6. In this case minimal total
oscillator quantum number is $E_{\mbox{\small min}}=2$. Hence 6 nucleon
system can be in the three ground state configurations $K_i$:

\begin{tabbing}
$K_1\equiv\left(00{1\over2}\right)^4\left(11{1\over2}\right)^2$:
\phantom{1234567} \= $JT =$ 01, 10, \\
$K_2\equiv
\left(00{1\over2}\right)^4\left(11{1\over2}\right)\left(11{3\over2}\right)$
:\>
$JT =$ 10, 11, 20, 21, \\
$K_3\equiv\left(00{1\over2}\right)^4\left(11{3\over2}\right)^2$: \>
$JT =$ 01, 10, 21, 30. \\
\end{tabbing}

Here $(elj)^n$ denotes the single shell. Displayed sequence of them
taken in ascending quantum number order stands for configuration.
According to the number of configurations with the same $JT$
values, situated at the right, we have to obtain:
one three-dimensional matrix with $JT=10$,
two two-dimensional matrices with $JT=01, 21$ and
three one-dimensional matrices with $JT=11, 20, 30$.
All shells and configurations are unambiguously characterized by their
$JT$ values, thus additional quantum numbers are not necessary.
The same specification scheme by $JT$ values will be valid and for
coefficients of expansion of intrinsic wave-function by shell model
functions. According to the Elliott and Skyrme theorem shell model states
characterized by $E_{\mbox{\small min}}$ contains the nuclear c.m.
in its ground state, hence
$a_{K{\mit\Delta};00,{\mit\Gamma}}^{EJ{\mit\Pi}T}=1$. In this case the
coefficients of expansion of intrinsic wave-function by shell model
functions are the CFP of intrinsic function.

Let us consider the intrinsic CFP characterized by $JT=30$.
This is the simplest case since they can be generated only in daughter
configuration $\left(00{1\over2}\right)^4\left(11{3\over2}\right)^2$.
Computational results are presented in Table 1.
\bigskip

Table 1

Intrinsic CFP of 6 nucleon system:

$E_{\mbox{\small min}}=2, \;
K=\left(00{1\over2}\right)^4\left(11{3\over2}\right)^2, \;
\Delta=1, \; JT=30$

\bigskip

\begin{tabular}{|c|c|c|c|c|c|c|} \hline
$\overline{\overline{K\Delta}}$ & $\overline{\overline{JT}}$ & 
$(el)_{-1}$ & $elsjt$ & state & $J''T''$ & CFP \\ \hline
$\left(00{1\over2}\right)^201\left(11{3\over2}\right)^230$ & 31 & 00 & 00001 &
$\raisebox{6pt}{\scriptsize 1}S_0$ &  01 & $\frac{1}{\sqrt{5}}$ \\
$\!\!\!\!\!\! 10$ & 20 &  & 00110 &
$\raisebox{6pt}{\scriptsize 3}S_1$ & 10 & $-\frac{1}{\sqrt{21}}$ \\
 & 30 &  &  &  & & $-\frac{1}{\sqrt{15}}$ \\
 & 40 &  &  &  & & $-\frac{3}{\sqrt{105}}$ \\ \hline
$\left(00{1\over2}\right)^3{1\over2}{1\over2}\left(11{3\over2}\right){3\over2}{1\over2}$ &
10 & 11 & 00110 &
$\raisebox{6pt}{\scriptsize 3}S_1$ & 20 & $-\frac{1}{2\sqrt{5}}$ \\
 & 11 & 00 & 11121 &
$\raisebox{6pt}{\scriptsize 3}P_2$ & 21 & $\frac{3}{2\sqrt{15}}$ \\
 & 20 &  & 11010 &
$\raisebox{6pt}{\scriptsize 1}P_1$ & 10 & $-\frac{1}{\sqrt{30}}$ \\
 &  & 11 & 00110 &
$\raisebox{6pt}{\scriptsize 3}S_1$ &  & $-\frac{1}{2\sqrt{15}}$ \\
 &  &  &  &
 & 20 & $-\frac{1}{\sqrt{30}}$ \\
 & 21 & 00 & 11111 &
$\raisebox{6pt}{\scriptsize 3}P_1$ & 11 & $\frac{1}{2\sqrt{5}}$ \\
 &  & 11 & 00001 &
$\raisebox{6pt}{\scriptsize 1}S_0$ &  & $\frac{1}{\sqrt{10}}$ \\
 &  & 00 & 11121 &
$\raisebox{6pt}{\scriptsize 3}P_2$ & 21 & $\frac{1}{\sqrt{10}}$ \\ \hline
$\left(00{1\over2}\right)^400$ &
00 & 00 & 22130 &
$\raisebox{6pt}{\scriptsize 3}D_3$ & 30 & $-\frac{1}{\sqrt{30}}$ \\
 &  & 22 & 00110 &
$\raisebox{6pt}{\scriptsize 3}S_1$ &  & $\frac{1}{\sqrt{30}}$ \\ \hline
\end{tabular}

\bigskip

Composition of grandparent configuration
$\overline{\overline{K{\mit\Delta}}}$
is clearly seen from displayed notations where every single shell is
specified by its total $JT$ values. Proposed in this paper set of quantum
numbers ${\mit\Gamma}J{\mit\Pi}TM_T$ for enumeration of antisymmetrical
states of A-particle system can be in natural way prolongated and for only
two particles. The corresponding set will be $lj\pi tm_t$. Here orbital
momentum $l$ plays the role of addition quantum number as ${\mit\Gamma}$,
since in the case of two particles $l$ is well defined quantum number.
So we follow the more usual spectroscopic notation
$\raisebox{6pt}{\scriptsize 2s+1}l_{\scriptstyle j}$
to denote the state of the subsystem containing two separated particles.
To avoid overloading the Table 1 we show only the values of the indices
which in turn were not repeated in the preceding row.

The obtained intrinsic CFP were used for intrinsic density matrices
calculations according the definition (2.31). As an example we present
intrinsic density matrix characterized by
$E=~E_{\mbox{\small min}}=2$, $JT=~10$
when the subsystem of two separated particles is in state
$\raisebox{6pt}{\scriptsize 3}S_1$

\begin{eqnarray}
  \begin{array}{c}
    {\bf W}^{2,10}(\,\raisebox{6pt}{\scriptsize 3}S_1) =
    \left(
      \begin{array}{rrr}
\frac{44}{135} & \frac{1}{135}  & \frac{-1}{27\sqrt{10}} \\
               & \frac{44}{135} & \frac{1}{27\sqrt{10}}  \\
               &                & \frac{17}{54}
      \end{array}
   \right)
  \end{array} \ .
\end{eqnarray}

Here we simplify notations of (2.31) and display in superscripts only
$E$ and $JT$.
${\bf W}^{2,10}(\,\raisebox{6pt}{\scriptsize 3}S_1)$ is symmetric matrix,
thus only upper triangle is shown.
It rows and the same columns are labelled by intrinsic CFP left hand bracet
indices. Namely rows labels are:

first -
$\Bigl(\left(00{1\over2}\right)^400,\left(11{1\over2}\right)^210\Bigr) 10$,

second -
$\Bigl(\Bigl(\left(00{1\over2}\right)^400,\left(11{1\over2}\right)
{1\over2}{1\over2}\Bigr){1\over2}{1\over2},
\left(11{3\over2}\right){3\over2}{1\over2}\Bigr) 10$,

third -
$\Bigl(\left(00{1\over2}\right)^400,\left(11{3\over2}\right)^210\Bigr) 10$.

Here notations of Table 1 are used except that
parentheses represents the momentum coupling order. Using the ordinary
shell model notations for intrinsic CFP is justified by
their clarity, origin of representation space construction procedures and
preservation of space dimension under transformation to Jacobi coordinates.
Last means that spaces labelled by ${\mit\Gamma}$ and $K{\mit\Delta}$
have equal dimensions.

\bigskip

Table 2

Diagonal elements of 6 nucleon system density matrices

\bigskip

\begin{tabular}{|c|ccc|c|cc|c|cc|c|} \hline
JT &   & 10    &       & 30    &    & 01 & 20 &    & 21 & 11 \\ \hline
$\frac{\mbox{Config.}}{\mbox{term}}$ &
$\left({1\over2}\right)^2$ \vline & $\left({1\over2}\right)\left({3\over2}\right)$
\vline & $\left({3\over2}\right)^2$ &
$\left({3\over2}\right)^2$ &
$\left({1\over2}\right)^2$ \vline & $\left({3\over2}\right)^2$ &
$\left({1\over2}\right)\left({3\over2}\right)$ &
$\left({1\over2}\left)\right({3\over2}\right)$ \vline & $\left({3\over2}\right)^2$ &
$\left({1\over2}\right)\left({3\over2}\right)$ \\ \hline
$\raisebox{6pt}{\scriptsize 3}S_1$ & $\frac{44}{135}$ & $\frac{44}{135}$ & $\frac{17}{54}$ &
$\frac{1}{3}$ & $\frac{3}{10}$ & $\frac{3}{10}$ & $\frac{1}{3}$ & $\frac{3}{10}$ & $\frac{3}{10}$ & $\frac{3}{10}$ \\ \hline
$\raisebox{6pt}{\scriptsize 3}S_1^{'}$ & $\frac{1}{810}$ & $\frac{8}{405}$ & $\frac{1}{81}$&
 & & & & & & \\ \hline
$\raisebox{6pt}{\scriptsize 1}S_0$ & $\frac{3}{10}$ & $\frac{3}{10}$ & $\frac{3}{10}$ &
$\frac{3}{10}$ & $\frac{14}{45}$ & $\frac{29}{90}$ & $\frac{3}{10}$ & $\frac{29}{90}$ & $\frac{14}{45}$ & $\frac{3}{10}$ \\ \hline
$\raisebox{6pt}{\scriptsize 1}S_0^{'}$ &  & & &
 & $\frac{1}{90}$ & $\frac{1}{45}$ & & & & \\ \hline
$\raisebox{6pt}{\scriptsize 1}P_1$ & $\frac{13}{270}$ & $\frac{13}{270}$ & $\frac{19}{270}$ &
$\frac{1}{30}$ & $\frac{1}{30}$ & $\frac{1}{30}$ & $\frac{1}{30}$ & $\frac{1}{30}$ & $\frac{1}{30}$ & $\frac{1}{30}$ \\ \hline
$\raisebox{6pt}{\scriptsize 3}P_0$ & $\frac{1}{10}$ & $\frac{1}{20}$ & &
& $\frac{1}{10}$ & & $\frac{1}{20}$ & $\frac{1}{20}$ & & $\frac{13}{180}$ \\ \hline
$\raisebox{6pt}{\scriptsize 3}P_1$ & $\frac{1}{5}$ & $\frac{1}{8}$ & $\frac{1}{20}$&
$\frac{1}{20}$ & $\frac{11}{45}$ & $\frac{13}{180}$ & $\frac{1}{8}$ & $\frac{47}{360}$ & $\frac{11}{180}$ & $\frac{17}{120}$ \\ \hline
$\raisebox{6pt}{\scriptsize 3}P_2$ &  & $\frac{1}{8}$ & $\frac{1}{4}$&
$\frac{1}{4}$ & & $\frac{1}{4}$ & $\frac{1}{8}$ & $\frac{17}{120}$ & $\frac{17}{60}$ & $\frac{11}{72}$ \\ \hline
$\raisebox{6pt}{\scriptsize 1}D_2$ & & & &
& & & & $\frac{1}{45}$ & $\frac{1}{90}$ & \\ \hline
$\raisebox{6pt}{\scriptsize 3}D_1$ & $\frac{2}{81}$ & $\frac{1}{162}$ & $\frac{1}{405}$&
& & & & & & \\ \hline
$\raisebox{6pt}{\scriptsize 3}D_2$ & & & &
& & & $\frac{1}{30}$ & & & \\ \hline
$\raisebox{6pt}{\scriptsize 3}D_3$ & & & &
$\frac{1}{30}$ & & & & & & \\ \hline
\end{tabular}

\bigskip

The empty places in the Table 2 means that there are any intrinsic
density matrices of such kind. Here
for the brevity in the configuration notations we don't display the first
closed shell $\left(00{1\over2}\right)^4$
and suppressed e and l in the higher single shells notations.

Presented results imply that even in $E_{\mbox{\small min}}$
approximation we can make same predictions concerning effects invoked
by tensor character of nuclear forces. The display of variety effects
arise from so-called mixing of states and have the same origin as in
deuteron case. According to the reduced Hamiltonian method \cite{5}
presence of bound states of A-particle system is explained by existence
of two-particle bound states of the so-called reduced Hamiltonian.
Direct numerical integration of reduced Hamiltonian equations with some
realistic nuclear interaction potential could supply corresponding
eigenvalues and percent of eigenstates mixing. For A = 6 that are only
$\raisebox{6pt}{\scriptsize 1}S_0$ and
$\raisebox{6pt}{\scriptsize 3}S_1-\raisebox{6pt}{\scriptsize 3}D_1$.
The admixture of $\raisebox{6pt}{\scriptsize 3}D_1$ term may consist
about 30\% . This admixture must be exhausted by corresponding terms
with higher orbital momentum in expansion of realistic density matrix
by intrinsic ones. The less part of admixture is exhausted in lover
energy approximations the more remain to the higher ones. Hence less
pronounced contribution of $\raisebox{6pt}{\scriptsize 3}D_1$ term
in $E_{\mbox{\small min}}$ approximation forces increasing influence
of admixed terms arising from higher energy approximations. Increasing
contribution of higher energy approximations in turn must increase
the calculated nuclear interaction radius.

As an example we consider the nuclei $\raisebox{6pt}{\scriptsize 6}Li$
and $\raisebox{6pt}{\scriptsize 6}He$, which ground states are
characterized by $JT=10$ and $JT=01$ correspondingly.
As can be seen from Table 2 the admixture of
$\raisebox{6pt}{\scriptsize 3}D_1$
term is absent for $\raisebox{6pt}{\scriptsize 6}He$ whereas it is not
the case for $\raisebox{6pt}{\scriptsize 6}Li$. Hence influence of higher
energy approximation must be more pronounced in the
$\raisebox{6pt}{\scriptsize 6}He$ case and as consequence it must have
the larger interaction radius, whitch is in full correspondence with
experimental results.

\section{CONCLUSIONS}
The only quantities really needed for calculation of identical particle
systems are two--particle density matrixes.
The developed formalism is particulary pointed to description of
energy spectrum formation mechanism and irregularities of root-mean-square
radii (Halo effect) in light exotic nuclei.
The method consistently lines the principles of antisymmetrization and
translational invariancy and is implicitly based on
Reduced Hamiltonian method.
The proposed density matrices are very suitable for exact realistic density
matrix expansion since it enable do not involve the
realistic spectator functions, devoted for description of the
remaining $A-2$ particle subsystem, into consideration.
It is well-known that such oscillator many particle functions form
slowly convergent series and thus the as large as possible set of states have
to be used.
This way the procedures of group theory for choosing
the 'best state' loose their sense.

The calculation procedures described above including ones for computation
of momentum recoupling matrices and Talmi-Moshinsky-Smirnov coefficients
were implemented in computer code.
The ordinary procedures for generating of complete many particle sets
for large spaces usually suffer of buildup of numerical errors.
Their presence could even lead to undeffiniteness of the obtained space
dimension.
Numerical instabilities do not concern the calculational procedures under
preposition.
All computations were performed with precise arithmetic using numbers
represented in the form $n/m/\sqrt{k}$, where $n$,$m$ and $k$ are integers.
The application of the precise arithmetic is possible due to the
orthogonal nature of the intrinsic density matrix formalism and the
particulary character of developed calculational procedures that involve
any numerical diagonalization and orthogonalization.
The effectiveness of the proposed approach could be illustrated by the
fact that intrinsic density matrices and related quantities of CFP nature
for energy up to $E_{\mbox{\small min}}+2$ and nuclei up to $A=11$ could
be calculated on 40 Mhz AT-386 within a few hours.



\begin{thebibliography}{99}
\bibitem{1}{\bf V.G. Neudatchin and Yu.F. Smirnov}, {\it Clusters of
nucleons in light nuclei} (Moscow, 1969).
\bibitem{2}{\bf I.V. Kurdyumov, Yu.F. Smirnov, K.V. Shitikova and
S.Kh.El.Samarai}, "Translationally invariant shell model,"
Nucl. Phys., A145, p. 593-612 (1970).
\bibitem{3}{\bf V. Vanagas}, {\it Algebraic Methods in Nuclear Theory}\,
(Vilnius, 1971).
\bibitem{4}{\bf G.P. Kamuntavi{\v c}ius}, "The reduced Hamiltonian method
in the theory of the identical particles systems bound states,"  Sov. J.
Particles and Nuclei, V. 20, No. 2, p. 261-292 (1989).
\bibitem{5}{\bf A. Deveikis and G. Kamuntavi{\v c}ius}, "Coefficients
of fractional parentage for nuclear shell model,"  Lithuanian Phys. J.,
V. 35, No. 1, p. 14-19 (1995).
\bibitem{6}{\bf A.J. Coleman}, "Structure of fermion density matrices,"
Rev. of Mod. Phys., V. 35, No. 3, p. 668-689 (1963).
\bibitem{7}{\bf G.P. Kamuntavi{\v c}ius}, "Simple functional-differential
equations for the bound-state wave-function components,"
Few-Body Systems, No. 1, p. 91-109 (1986).
\bibitem{8}{\bf G. P. Kamuntavi{\v c}ius}, "Coefficients of fractional parentage
of microscopic atomic nucleus models,"  Lithuanian
Phys. Collection., V. 28, No. 2, p. 135-147 (1988).
\bibitem{9}{\bf J.Levinsonas}, "Konfig{\= u}racijos, turin{\v c}ios
kelet{\c a} sluoksni{\c u}, kilminiai koeficientai,"
LTSR Moksl{\c u} Akademijos Darbai V. 4, No. 4, p. 17-31 (1957).
\bibitem{10}{\bf L. Trlifaj}, "Simple formula for the general
oscillator brackets,"
Phys. Rev. C, V. 5, No. 5, p. 1534-1539 (1972).
\bibitem{11}{\bf J. Dobe{\v s}}, "A new expression for harmonic
oscillator brackets,"
J. Phys. A.: Math.Gen., V. 10, No. 12, p. 2053-2059 (1977).
\end{thebibliography}
\end{document}